\def\half{\frac{1}{2}}

\def\rect{{\rm rect}}
\def\half{{\frac{1}{2}}}
\def\imag{{\rm Im}} 
\def\conv{{\ast}}
\def\wigmap{{\mathcal W}}
\def\det{{\rm det}}
\def\moyalstar{\star}
\def\Brill{{\mathcal B}}

\documentclass[%
 reprint,
nofootinbib,
 amsmath,amssymb,
 aps,
prstab,
]{revtex4-1}

\usepackage{xcolor}

\usepackage{placeins}
\usepackage{graphicx}% Include figure files
\usepackage{dcolumn}% Align table columns on decimal point
\usepackage{bm}% bold math
\usepackage{hyperref}% add hypertext capabilities
\usepackage{natbib}
\DeclareMathOperator\erf{erf}

\begin{document}

\title{Propagation of partially coherent radiation using Wigner functions}

\author{Boaz Nash}%
 \email{bnash@radiasoft.net}
 \homepage{}
\author{Nicholas Goldring}
\author{Jonathan Edelen}
\author{Stephen Webb}
\affiliation{RadiaSoft LLC, 3380 Mitchell Lane, Boulder CO 80301}
 \altaffiliation[]{}%Lines break automatically or can be forced with \\
 
 \author{Rafael Celestre}
 \affiliation{ESRF - The European Synchrotron}

\date{\today}% It is always \today, today,

\begin{abstract}
Undulator radiation from synchrotron light sources must be transported down a beamline from the source to the sample. A partially coherent photon beam may be represented in phase space
using a Wigner function, and its transport may use some similar techniques as is familiar in particle beam
transport. We describe this process in the case that the beamline is composed of linear focusing and defocusing
sections as well as apertures. We present
a compact representation of the beamline map involving linear transformations and convolutions.
We create a 1:1 imaging system (4f system) with a single slit on the image plane and observe the radiation downstream to it. We propagate a Gaussian beam and undulator radiation down this sample beamline,
drawing parameters from current and future ultra low emittance light sources. We derive an analytic expression for the partially coherent Gaussian case including passage through a single slit aperture.
We benchmark the Wigner function calculation against the analytical expression and 
a partially coherent calculation in the Synchrotron Radiation Workshop (SRW) code.
\end{abstract}

\pacs{Valid PACS appear here}% PACS, the Physics and Astronomy
\maketitle

%\tableofcontents

\section{Introduction}
%% !TEX root = ../beamlineMapPrab.tex

Synchrotron radiation sources, either storage ring or FEL-based, require optical beamlines to transport the radiation to the experimental sample. As performance of these sources is being pushed to lower emittance and higher coherence, the beamline performance and modeling must be accordingly improved ~\cite{SanchezdelRio2019Nov}.

Models of radiation transport through the beamline elements exist in a hierarchy of levels of accuracy and complexity. At the simplest level, one can use analytical formulae to propagate beam sizes, divergences and coherence lengths through an idealized beamline (see ~\cite{Shi2017Aug}). At the next level is the geometric optics description using a ray-tracing approach (e.g. SHADOW~\cite{Rio:2011aa}).  At a higher level of complexity, one may use a physical optics approach which requires wavefront propagation software (e.g. SRW~\cite{Chubar:1998vi}). The wavefront propagation allows the inclusion of diffraction effects in coherent optics, but is more
 computationally intensive than the ray-tracing approach.  
 
 To go beyond fully coherent optics, accurate representation of radiation requires the model include partial coherence resulting from considerations of statistical optics \cite{Goodman2015May, Geloni2008Apr}. For synchrotron radiation, this partial coherence results from the finite electron beam size causing randomness in the phase of the emitted radiation. In the wavefront propagation method, this may be taken into account by propagating multiple initial wavefronts, either via a sampling of the phase space of initial electron beam~\cite{Chubar2011Sep} or by a coherent mode decomposition~\cite{Glass2017Oct,Li2020Aug,Coisson1997Sep}.
 
Another approach to treating partially coherent radiation involves the use of Wigner functions. The Wigner function formalism for synchrotron radiation was pioneered by K.-J.~Kim~\cite{KimWigner1986}.  The Wigner function was originally developed in quantum statistical mechanics ~\cite{Curtright,Curtright2011Apr, Wigner1932Jun}, as an alternative phase-space representation to the density operator. In an optical context, the Wigner function may represent the types of systems described in statistical optics, where multiple wavefronts are simultaneously
present with random phase relations between them. The properties of Wigner functions and the relation between the quantum mechanics and optics context are described by Bazarov et al.~\cite{Bazarov}.
 
Although Wigner functions for fully and partially coherent synchrotron radiation have been computed, they have not been widely used in the propagation down beamlines.
In this paper, we demonstrate propagation of the fully and partially coherent Wigner functions through a simplified beamline.  
We compute beamline maps that may be applied to any initial condition Wigner Function without having to be recomputed. We limit ourselves to linear maps, (so-called ABCD matrix, in the optics literature~\cite{Siegman1986, Ferrero2008Aug}) under which the Wigner function transforms in a straightforward manner. We will also consider physical apertures, focusing on the case of single slits, an important element in most 
x-ray beamlines. 
We assume separable radiation such that we may work with 2D Wigner functions~\cite{Bazarov, Tanaka2014Jun}. The extension to higher dimensionality is straightforward. Including non-linear elements, such as spherical aberrations in a lens, will be a topic of future work. 

As the size of the electron beam increases, the Wigner function becomes dominated by the Gaussian electron beam and the coherence decreases. To understand this transition, we consider
Gaussian initial Wigner functions. We are able to derive an analytic expression for the diffraction of a partially coherent Gaussian through a single slit. We may thus validate our
algorithms for linear transport and passage through an aperture. In addition, we may compare the analytical Gaussian result to the case of undulator radiation and gain understanding as to when the Gaussian result may be adequate in simulation.  Finally, we validate our undulator radiation transport using our simplified Wigner function map method to a partially coherent SRW calculation.

\section{X-ray Beamline Modeling}

%% !TEX root = ../beamlineMapPrab.tex

The goal of X-ray beamline modeling is to take an initial radiation field and propagate that field through a complex series of elements to accurately predict what the wavefront will look like when it reaches the sample.  There are several methods of approaching this: we survey physical optics in the Appendix \ref{app:cwo}, for example.  We start with a summary of the properties of the radiation Wigner function.  Next, we provide a brief discussion of linear geometric optics including apertures.  Finally, we show how to evolve the Wigner function under the action of a matrix-aperture beamline. 

Consider an optical ray and attribute to it a wavelength $\lambda$ following a trajectory starting at position $s=0$ where the radiation is created and ending at $S=L$ at the end of the beamline. This trajectory will in general not be straight due to reflections off of mirrors and gratings and passing through other optical elements. At each position $s$, along the trajectory,
we assign transverse phase space coordinates, $\vec{z}$,
\begin{equation}
\vec{z} = \begin{pmatrix}
x\\
\theta_x\\
y\\
\theta_y
\end{pmatrix}
\end{equation}
where $x$ and $y$ are transverse coordinates, $\theta_x$ and $\theta_y$ are corresponding angles with $\theta_x = \frac{dx}{ds}$ and $\theta_y = \frac{dy}{ds}$.
 \footnote{Note that we will ignore the longitudinal phase space coordinates in this work. A 6D phase space treatment would include the longitudinal position and relative wavelength deviation. In particular we ignore pulse length effects and assume a monochromatic
 beam.}

\subsection{Radiation Wigner functions}

We represent the radiation along the trajectory by means of a Wigner function $W(\vec z)$. Many of the properties of Wigner functions have been reviewed by Bazarov~\cite{Bazarov}. We mention several of them so that our treatment here is self-contained. First, the Wigner
function is normalized:
\begin{equation}
\int d\vec z~ W(\vec z) = 1
\end{equation}
We adopt this normalization for clarity of presentation and close connection to the corresponding quantum mechanical formalism. The Wigner function is related to the brightness (or brilliance) function $\Brill$ by means of an overall factor of the radiation flux $\phi$:
\begin{equation}\label{BrillWigConnection}
\Brill(\vec z;s) = \phi(s) W(\vec z;s)
\end{equation}
Thus, as the radiation moves along the beamline, progressing in $s$ and passes through absorbing elements (such as apertures that we consider here), the flux $\phi$ will reduce, but the normalization of $W(\vec z;s)$ will remain constant.

Now, suppose we know the Wigner function at $s=0$, $W_0(x,\theta_x,y,\theta_y)$. We will assume that $W_0$ is separable; that is that $W_0$ obeys\footnote{Note that for the case of undulator radiation, this condition only approximately holds.  On resonance, the condition is satisfied to within a few percent over a wide range of emittance values.  See Fig. 6 in \cite{Tanaka2014Jun}.}
\begin{equation}
W_0(x,\theta_x,y,\theta_y) = W_x(x,\theta_x) W_y(y,\theta_y)
\end{equation}

In the examples we consider, the Wigner function remains separable throughout the beamline, and thus we may consider propagation of the components
separately. We thus refer to $W_x(x,\theta_x)$ or $W_y(y,\theta_y)$ simply by $W(x,\theta)$.

Now, suppose that $W(x,\theta)$ represents fully coherent radiation. Then, there exists an electric field $E(x)$ such that 
\begin{equation}\label{wignerEqn01}
W(x,\theta) = \frac{1}{\lambda}  \int_{-\infty}^{\infty} E^* \left(x-\frac{\phi}{2}\right)E\left(x+\frac{\phi}{2}\right)e^{-\frac{2\pi i}{\lambda}\phi \theta}d\phi.
\end{equation}
We may write this equation in an operator form:
\begin{equation}
W(x,\theta) = \wigmap(E(x))
\end{equation}
where we refer to $\wigmap$ as the Wigner transform operator, or $W(x,\theta)$ as the Wigner function associated with $E(x)$.
In the case of fully coherent radiation, the electric field may be reconstructed from the Wigner function as follows~\cite{Bazarov}
\begin{equation}\label{WtoE}
E^*(x)E(0) = \frac{1}{\lambda} \int_{-\infty}^{\infty} W \left(\frac{x}{2},\theta \right) e^{\frac{2 \pi i}{\lambda} x \theta} d\theta.
\end{equation} 

Now, in the case where $W$ is partially coherent, there does not exist a single, well-defined wavefront associated with the Wigner function.
Rather, there exists a whole sequence of fields $E_j(x)$ with $j=1\dots \infty$. The Wigner function is given as the (infinite) sum of the Wigner
transforms associated with the $E_j$:
\begin{equation}\label{coherentModeDecomp}
W(x,\theta) = \sum_j \wigmap(E_j(x))
\end{equation}
The decomposition of a given partially coherent Wigner function into a set of underlying modes $E_j$ is not generally unique. The process of finding
an astute choice of such modes is known as ``coherent mode decomposition.''  The modes are often taken to be orthornormal (see section 4.7 in \cite{Mandel1995Sep}).  However, in the case of synchrotron radiation, one may also consider the single electron modes as a form of coherent mode decomposition also satisfying Eqn. (\ref{coherentModeDecomp}).

From the Wigner function, one may compute the quantity, $\mu$, known as the degree of coherence via the following expression.

\begin{equation}\label{degCoherence}
\mu^2 = \lambda \int W^2(x,\theta) dx d\theta
\end{equation}

In the fully coherent case where $W$ may be derived from an electric field, we find $\mu = 1$.  For the partially coherent case, $\mu < 1$.

\subsection{Linear Geometric Optics}

Along an X-ray beamline, there are optical elements which the radiation will interact with. We  include these in our model by including a varying optical path difference, that is, a phase and amplitude modulation, as a function of the Cartesian coordinates $(x,y;s)$. This optical path length difference is often a function of the index of refraction $n(x,y;s)$ for transmission elements. In addition we will include physical apertures, which allow radiation within a certain transverse
region to pass unimpeded, and absorb all radiation outside of that region. The aperture elements may be described by transfer functions $t(x,y)$ which describe the region where rays may pass, and where they are absorbed. In fact, we may allow more general aperture
elements with values between 0 and 1 in which the intensity of the ray may be reduced but not fully absorbed.

Now consider another ray, starting at a different initial condition $\vec Z_0$. The evolution of this ray down the beamline may be described by the action of the following Hamiltonian
\begin{equation}
H(x,\theta_x,y,\theta_y;s) = -\sqrt{n^2(x,y;s) - \theta_x^2 - \theta_y^2}.
\end{equation}
with $\theta_x$ and $\theta_y$ playing the role of momenta, and with position along the trajectory $s$ as independent variable. $n(x,y;s)$ is the local index of refraction that the radiation is passing through. The result of this Hamiltonian formulation for 
geometric optics is that the offset ray will follow Hamilton's equations:
\begin{equation}
\dot Z_i = J_{ij}\frac{\partial H}{\partial Z_j}
\end{equation}
with summation over the repeated index $j$ implied and the dot representing $\frac{d}{ds}$. For 4D phase space (in the case that we ignore variation in the $z$ and $\delta$ coordinates), the matrix $J$ is given by
\begin{equation}
J = \begin{pmatrix}\label{JMAT}
0 & 1 & 0 & 0\\
-1 & 0 & 0 & 0\\
0 & 0 & 0 & 1\\
0 & 0 & -1 & 0
\end{pmatrix}
\end{equation}

If we allow arbitrary index of refraction $n(x,y)$ in our model beamline, then the equations of motion will be non-linear. For the purposes of this paper, we will restrict to the approximation that the index of refraction varies quadratically, leading to
linear equations of motion for the ray tracing. In particular, as our model for the index of refraction $n(x,y;s)$, we will assume it to be constant along the optical axis and then to fall off quadratically in the transverse directions. Thus, we parametrize it as follows \footnote{Note that we leave out a coupling term in the Hamiltonian proportional to $xy$ so that the separable condition is satisfied.} :
\begin{equation}
n(x,y;s) = n_0(s) -\kappa_x(s)  x^2 -\kappa_y(s) y^2.
\end{equation}
Expanding the Hamiltonian for small angles, to quadratic order, we find
\begin{equation}
H(x,y,\theta_x,\theta_y;s) \approx -n_0(s) + \frac{\theta_x^2 + \theta_y^2}{2 n_0} +\kappa_x(s) x^2 +\kappa_y(s) y^2.
\end{equation}
As mentioned, this Hamiltonian will lead to linear equations of motion. The solution may thus be expressed in matrix form as
\begin{equation}\label{TransferMatrix}
\vec Z(s) = M(s) \vec Z_0
\end{equation}
One may solve these equations and produce a linear map for a given beamline section.  All the optical elements (excluding the apertures which we deal with separately are thus captured in the transfer matrix $M(s)$ varying along the beamline.
Because the ray tracing dynamics are derived from a Hamiltonian, we are assured that the resulting transfer matrix is symplectic. That is:
\begin{equation}
M(s)^T J M(s) = J
\end{equation}
for all $s$ along the beamline. The matrix $J$ is given in Eqn. (\ref{JMAT}).

Although we have formulated this section in terms of a Hamiltonian theory to bring out some of the formal properties of the propagation, the transfer matrix may also be computed for realistic beamlines using ray tracing software. See \cite{Nash2020Aug}
for an example of this calculation for a KB mirror system.

In the next section, we describe the evolution of the Wigner function under the matrix $M$. This covers both the fully coherent and partially coherent case. In the fully coherent case, a different formalism is possible for the propagation: that of the linear canonical
transform (LCT). We outline this in Appendix A.

\subsection{Partially coherent propagation with Wigner function}

The evolution equation for the Wigner function is given as follows~\cite{Curtright2011Apr}
\begin{equation}
\frac{\partial W(x,\theta_x, y, \theta_y;s)}{\partial s} = [W,H]_\moyalstar,	
\end{equation}
where the Moyal bracket is defined for arbitrary phase space functions $f$ and $g$ as
\begin{equation}
[f,g]_{\moyalstar} = \frac{1}{i \lambdabar}(f\moyalstar g-g\moyalstar f)
\end{equation}
and the Moyal star is given by
\begin{equation}
\moyalstar = e^{\frac{i\lambdabar}{2} \left(\overleftarrow\partial_x \overrightarrow\partial_{\theta} - \overleftarrow\partial_{\theta} \overrightarrow\partial_x \right)}
\end{equation}
with the arrows representing action of the derivative, either to the left or right, depending on arrow orientation.

Fortunately, in the case of a quadratic Hamiltonian, evolution of the Wigner function is much simpler and more intuitive.  The Moyal bracket reduces to the Poisson bracket giving classical evolution (again using the quantum/classical mechanics analogy). One finds that the motion in phase space is a linear transformation. These considerations allow us to formulate our approach.   In particular, consider a beamline where the geometric optics is defined by a transfer matrix $M$ acting on the phase space vector $\vec{z}$:
\begin{equation}
\vec{z}_f=M \vec{z}_i
\end{equation}
The Wigner function evolves along this beamline according to
\begin{equation}
W_f(\vec{z}) = W_i(M\vec{z}).
\end{equation}

We may describe this transformation with the operator, $U_M$, defined as
\begin{equation}\label{UMoperator}
U_M(W(\vec{z})) = W(M\vec{z})
\end{equation}

By performing a change of variables, one may show that the degree of coherence $\mu$ is conserved under linear transport $U_M$. That is,
\begin{equation}\label{degCoherenceLinear}
\mu(W(\vec z)) = \mu(W(M\vec z))
\end{equation}
The degree of coherence is not conserved after passing through an aperture, which we now describe.

We would now like to consider the way in which Wigner functions are impacted by physical apertures. As described by Bazarov, for the electric field, the effect of the aperture is given by
\begin{equation}
E_{s'}(\vec{x}) = E_{s}(\vec{x})t(\vec{x})
\end{equation}
where $t(\vec{x})$ is the transmission function of the aperture.

In terms of the Wigner function, the action of the aperture is given by the partial convolution (in the angular variable) of the aperture Wigner function:

\begin{equation}\label{APoperator}
W_{s'} = W_{s} \conv_{\theta} W_{t} \equiv \mathcal{A}_{t}W_{s} 
\end{equation}

The aperture Wigner function $W_{t}$ is given by applying the Wigner transform to the aperture transmission function. That is, we apply Eqn (\ref{wignerEqn01}) where the aperture transmission function $t(x)$ plays the role of the electric field.  

Since an aperture results in absorption of radiation, the normalization of the Wigner function would change after passing through.  As given in Eqn. (\ref{BrillWigConnection}), the Wigner function is related to the brightness by a factor of the total flux.  For simplicity, we will ignore the changing value of flux along the beamline and consider Wigner function to be consistently normalized throughout.  Thus we normalize the transmission function according to Eqn. (\ref{enormcoord}) and thus $\wigmap(t_j)$ will be normalized according to Eqn. (\ref{wignerNorm01}). Because of the different
normalization, the Wigner transform of a transmission function could be considered as a sort of filter that acts on the incoming radiation Wigner Distribution Function (WDF). Thus, part of the work of propagation of WDFs through beamlines including apertures involves the calculation of the
Wigner filter functions. We will give an example of this function for the single slit aperture in a later section. A substantial part of the understanding about diffraction effects can be gleaned by examination of these Wigner filter functions.

\subsection{Matrix-aperture beamlines}

\begin{figure*}[!tbh]
	\centering
{\includegraphics[width=\textwidth,height=3.2in,keepaspectratio]{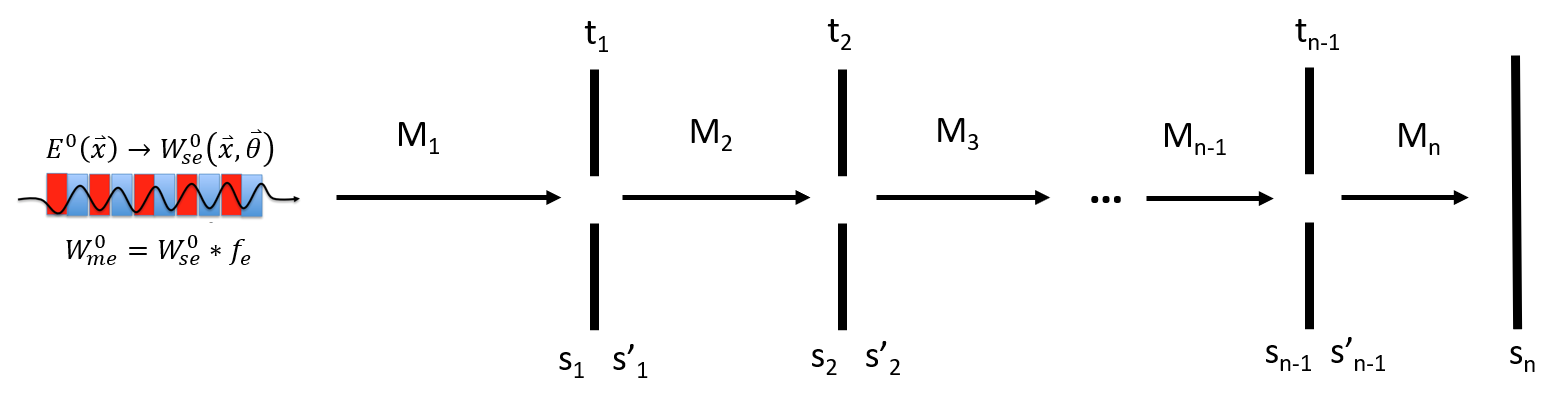}}
\caption{Matrix aperture beamline schematic.}
\label{beamlineSchematic01}
\end{figure*}

Consider the beamline schematic as shown in Fig. \ref{beamlineSchematic01} consisting of an undulator source with an electron beam and subsequent sections which may be described by matrices, $M_j$, and apertures with transmission function, $t_j(x)$.  An electron beam with distribution $f_e(\vec{z})$ passes through an undulator producing synchrotron radiation.  Let $E^0(\vec{x})$ be the electric field produced by a single electron as it appears at the center of the undulator.  We may now construct the multi-electron Wigner function for the undulator radiation as will be described 
in Section IV-A (see Eqn. (\ref{WelecDist})).

%From $E^0(\vec{x})$, we construct the single electron Wigner function, $W^0_{se}(\vec{x},\vec{\theta})$ according to Eqn. \ref{wignerEqn01}.  The multiple electron Wigner function, $W^0_{me}(\vec{x},\vec{\theta})$, can then be computed by the convolution

We consider an axial ray coming from the center of the undulator and proceeding along the beamline until the final observation plane located at position $s_n$.  We note that the optical axis described by this axial ray is not necessarily a straight line.  In particular, mirrors will cause angular deviations from the central trajectory and at each point, the transverse coordinates are relative to the direction of the central ray.  Along the beamline, there are apertures located at positions $s_1$ through $s_{n-1}$ which are represented by transmission functions $t_1$ through $t_{n-1}$.  For the purpose of our simplified beamline, non-linear aberrations will be ignored and we will assume that the transport between apertures $j-1$ and $j$ may be represented by the matrix $M_j$.  

Following Eqn. (\ref{UMoperator}), we find an operator for this beamline section given by $U_{M_j}$ that simply transports the phase space by applying the matrix $M_j$.  We have thus defined the operator for propagation through sections of linear transport, $U_{M_j}$, that may include mirrors, lenses, and other elements when remaining close to the optical axis.  Likewise, the apertures may also be represented by operators, $\mathcal{A}_{t_j}$, as given by Eqn. (\ref{APoperator}).  The operator for the entire simplified beamline may then be given by

%\begin{equation}
%W_{j}(\vec{z}) = U_{M_j}W_{j-1}(\vec{z}) = W_{j-1}(M_j\vec{z})
%\end{equation}

%The action of the aperture on an electric field is simply given by

%\begin{equation}
%E_{s'_j}(\vec{x}) = E_{s_j}(\vec{x})t(\vec{x})
%\end{equation}

%In terms of the Wigner function, the action of the aperture is given by the partial convolution of the aperture Wigner function:

%\begin{equation}
%W_{s'_j} = W_{s_j} \star_{\theta} W_{t_j} \equiv \mathcal{A}_{t_j}W_{s_j} 
%\end{equation}

%The aperture Wigner function is given by applying Eqn. \ref{wignerEqn01} where the aperture transmission function plays the role of the electric field.  Note that we normalize the transmission function according to Eqn. \ref{enormcoord} and thus $W_{t_j}$ will be normalized according to Eqn. \ref{wignerNorm01}.  

\begin{equation}
\mathcal{O}_{BL} = U_{M_n}\mathcal{A}_{t_{n-1}}U_{M_{n-1}}...\mathcal{A}_{t_2}U_{M_2}\mathcal{A}_{t_1}U_{M_1}
\end{equation}

Then the fully coherent and partially coherent simulations can be simply written as 

\begin{eqnarray}
W_{se,n} &=& \mathcal{O}_{BL}W^0_{se}\\
W_{me,n} &=& \mathcal{O}_{BL}W^0_{me}
\end{eqnarray}

respectively.

\section{Examples -- Gaussian radiation}
%% !TEX root = ../beamlineMapPrab.tex

\begin{figure*}[!tbh]
	\centering
{\includegraphics[width=\textwidth,height=1.5in,keepaspectratio]{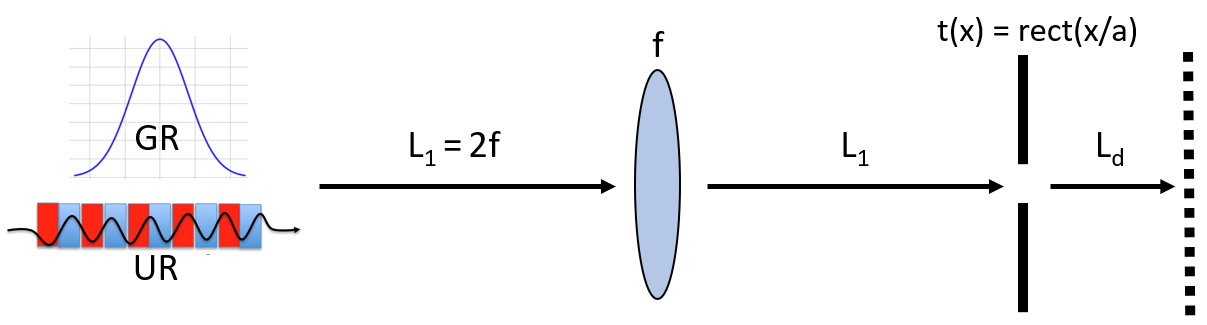}}
\caption{Simple matrix-aperture beamline example.}
\label{beamlineSchematic02}
\end{figure*}

We consider Gaussian radiation (GR) and undulator radiation (UR) propagating through a simple matrix-aperture beamline doing a 1:1 imaging of the source with a horizontal slit at the image plane and observing the radiaion downstream to it.  A schematic for this beamline is depicted in Figure \ref{beamlineSchematic02} and the corresponding source and beamline parameters are provided in Table \ref{paramTable01}.

\begin{table}[!htbp]
\caption {Numerical Simulation Parameters} \label{paramTable01} 
\begin{tabular}{lcc}
\hline
\multicolumn{3}{l}{\textbf{Undulator radiation}} \\ \hline
Length, $L_u$                           & 2.31                    & m                   \\
Period, $\lambda_u$                    & 0.033                     & m                    \\
Max. field, $B$                            & 0.7                     & T                    \\
Deflection parameter, $k$                            & 2.157                     &                     \\
First resonant energy, $E_1$                           & 3.115                     & keV                    \\
Wavelength, $\lambda_1$                           & 3.98                    & \AA                   \\ \hline
\multicolumn{3}{l}{\textbf{Electron beam (APS-U)}}                   \\ \hline
Energy, $E$                            & 6.0                     & GeV                    \\
Current, $I$                            & 200                     & mA                    \\
Horizontal emittance, $\epsilon_x$                       & 42.2                     & pm                    \\
Horizontal RMS size, $\sigma_x$                       & 14.44                     & $\mu$m                    \\
Horizontal RMS divergence, $\sigma_{x'}$                      & 2.92                     & $\mu$rad                     \\
Vertical emittance, $\epsilon_y$                       & 4.20                     & pm                    \\
Vertical RMS size, $\sigma_y$                       & 2.82                     & $\mu$m                    \\
Vertical RMS divergence, $\sigma_{y'}$                      & 1.49                     & $\mu$rad                    \\ \hline
 \multicolumn{3}{l}{\textbf{Beamline parameters}} \\ \hline
Distance to lens, $L_1$                    & 30.0                     & m                    \\
Lens focal length, $f$                            & 15.0                     & m                    \\
Aperture width, a                            & 16.7                     & $\mu$m                    \\
Final drift length, $L_d$                           & 0.1                    & m                   \\ \hline                  
\end{tabular}
\end{table}

We first illustrate many of the useful properties of radiation Wigner Function Distribution (WFD) propagation with the example of Gaussian radiation.  The one-to-one imaging section preserves the coherence properties\footnote{This assumes that that there is no beam cropping and that the imaging system can resolve the radiation source.} and thus the first element to consider is the single slit aperture.  Propagation through the aperture and subsequent propagation through free space allows us to observe the impact of decreasing coherence on the WFD and corresponding intensity pattern.  As the divergence becomes sufficiently large, the oscillations in $\theta$ in the aperture Wigner filter are washed out, and coherent diffraction effects are seen to be destroyed.  The GR case has the advantage of exploiting the analytical expressions previously derived for benchmarking our numerical WFD transport methods.

We then explore the more complex case of UR propagating through this beamline.  Changing the electron beam emittance allows us to control the degree of partial coherence.  At small electron beam emittances, the WFD is dominated by the single electron WFD, but as the electron beam emittance increases, this structure becomes less relevant and the WFD asymptotically reduces to the GR case.  In order to confirm the accuracy of our calculations, we have set up the same beamline model in SRW.  Due to the separability of the horizontal and vertical components, we are able to demonstrate a direct comparison between SRW and our own partially coherent WFD based propagation results. 

Although our WFD based methods are substantially less time-consuming than full multi-electron partially coherent calculations, significant computational issues remain present regarding memory and runtime.  Due to the complexities in the radiation and aperture WFD patterns, sufficient resolution is required to obtain accurate results.  Often times this runs into conflict with memory and runtime requirements for the calculations.  For this reason, we report the grid sizes ($n_x$ number of points in position and $n_{\theta}$ points in angle) used in each numerical calculation.

We first consider a Gaussian beam propagating through the simple matrix aperture beamline where we can compare our numerical results to the analytical expression given by Eqn. (\ref{singleSlitGsnAnalytic}). 

\subsection{Analytic calculation for Gaussian radiation with single slit aperture}\label{analyticgauss}

In this section, we derive an analytic expression for the propagation of a Gaussian Wigner function through a single slit aperture of width $a$.  We are able to find an analytic expression for the radiation immediately after the slit as well as the radiation after having drifted some distance beyond.  The aperture is described by the transmission function
\begin{equation}
t(x) = \rect\left(\frac{x}{a}\right)
\end{equation}
with
\begin{equation}
\rect(x) = 
\begin{cases} 
1, & |x|< \half \\
\frac{1}{2} & |x| = \half \\
0, & |x| > \half
\end{cases}
\end{equation}

The corresponding Wigner function is given by \cite{Roman-Moreno2003Jan}
\begin{eqnarray}\label{Wss}
W_{ss}(x,\theta) &=& \rect\left(\frac{x}{a}\right) \frac{2\sin\left[ \frac{2\pi \theta}{\lambda} (-2|x| + a)\right]}{\theta} \\
&\equiv& \frac{2 \sin( \theta Q)}{\theta} \\
Q & =&  2\pi \frac{a - 2|x|}{\lambda}
\end{eqnarray}

\begin{figure*}[!tbh]
	\centering
{\includegraphics[width=\textwidth,height=2.2in,keepaspectratio]{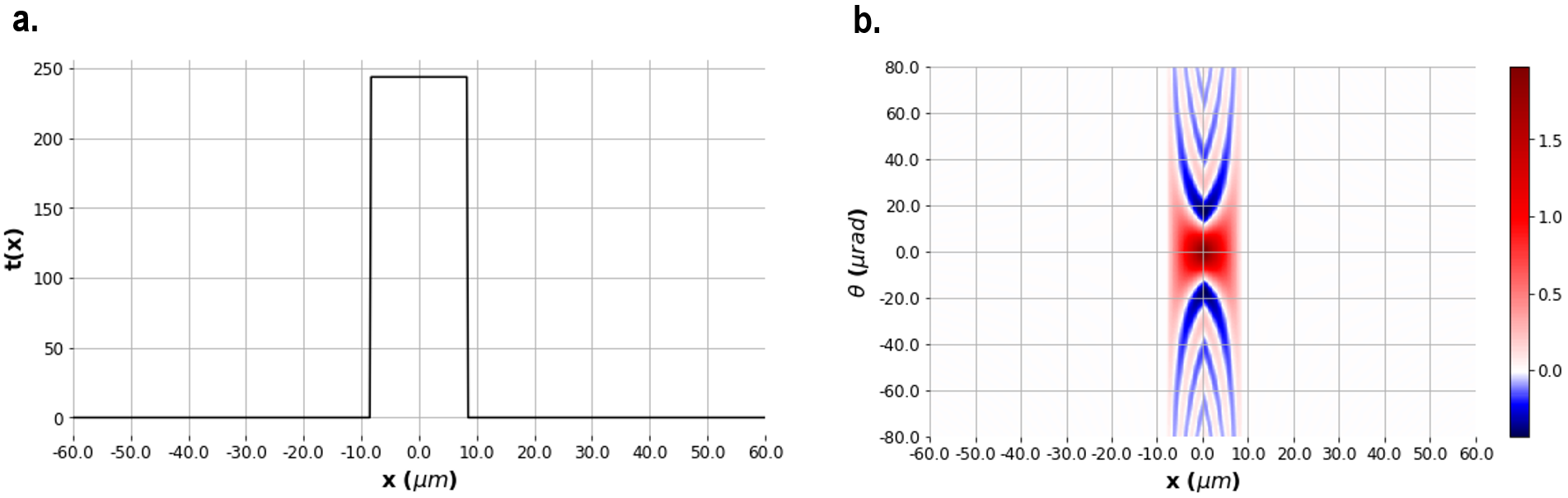}}
\caption{Single slit aperture (a) transmission function and (b) corresponding Wigner filter function.}
\label{sftransmission_sfwigner}
\end{figure*}

Plots of the single slit aperture transmission function and corresponding WFD are displayed in Figure \ref{sftransmission_sfwigner}.

Let us consider a partially coherent Gaussian Wigner function given by
\begin{equation}\label{wignerGsn}
W(x,\theta) = \frac{1}{2\pi \sigma_x \sigma_\theta} e^{-\frac{x^2}{2\sigma_x^2} - \frac{\theta^2}{2\sigma_\theta^2}}
\end{equation}
Let us write the relationship between the position and angular spreads as
\begin{equation}\label{diffracLimEqn}
\sigma_x \sigma_\theta = m^2\frac{\lambda}{4\pi}
\end{equation}
where $m^2$ is the beam quality factor known in the laser optics literature\footnote{The beam quality factor is typically represented by a capital $M$.  We have used a lower-case $m$ here so as not to confuse this quantity with the transfer matrix, $M$}.  In the case that $m^2=1$, the Wigner function represents a coherent wavefront\footnote{We note that for non-Gaussian wavefronts, in the coherent case, the product $\sigma_x \sigma_\theta$ exceeds $\frac{\lambda}{4\pi}$. For undulator radiation, one has approximately  $\frac{\lambda}{2\pi}$.  See e.g. \cite{Walker2019May,Khubbutdinov2019Nov,Nash2019Jan,Nash2019Sep} .}. For $m^2>1$, this Wigner function represents partially coherent radiation.

To propagate the Wigner function through the slit, we perform a convolution of the Gaussian with Eqn. (\ref{Wss}) in the $\theta$ variable.
That is:
\begin{eqnarray}
W_{s_1'} &=& W_{s_1} \conv_{\theta} W_{ss}\\ 
&=& \rect\left(\frac{x}{a}\right) \frac{2}{2\pi \sigma_x\sigma_\theta} e^{-\frac{x^2}{2\sigma_x^2}} \mathcal{I}(\theta,Q)
\end{eqnarray}
with
\begin{equation}
\mathcal{I}(\theta,Q) =  \int_{-\infty}^{\infty} e^{-\frac{(\theta - \tau)^2}{2\sigma_\theta^2}} \frac{\sin(\tau Q)}{\tau} d\tau
\end{equation}
This integral may be performed (see Appendix D for details) with the result
\begin{equation}
\mathcal{I}(\theta,Q) = \pi e^{-\frac{\hat\theta^2}{2}} \imag\Bigg\{ i \erf\left(\frac{\hat Q + i \hat \theta}{\sqrt{2}}\right) \Bigg\} 
\end{equation}
where $\imag$ represents taking the imaginary part of the argument and $\hat Q = Q\sigma_\theta$ and $\hat \theta = \theta/\sigma_\theta$. Writing it all out explicitly
to see the $x$ and $\theta$ dependence, and adding the drift following the aperture, we have:

\begin{widetext}
\begin{equation}\label{singleSlitGsnAnalytic}
W_{s_1'}(x,\theta) =   \frac{1}{ \sigma_x\sigma_\theta} \rect\left(\frac{x-L_d\theta}{a}\right)e^{-\frac{(x-L_d\theta)^2}{2\sigma_x^2}-\frac{\theta^2}{2\sigma_\theta^2}}  \imag\Bigg\{i \erf\left(\frac{(a-2|x-L_d\theta|)\frac{\sigma_\theta}{\lambda} + i \frac{\theta}{\sigma_\theta}}{\sqrt{2}}\right) \Bigg\}
\end{equation}
\end{widetext}
where $L_d$ is the drift length following the single slit aperture.

From Eqn. (\ref{degCoherence}), and using the Gaussian Eqn. (\ref{wignerGsn}), we find the expression for degree of coherence, $\mu$, before passing through the slit to be

\begin{equation}\label{degCoherenceExpr}
\mu = \frac{1}{m}
\end{equation} 

\begin{figure*}[!tbh]
	\centering
{\includegraphics[width=\textwidth,height=3.2in,keepaspectratio]{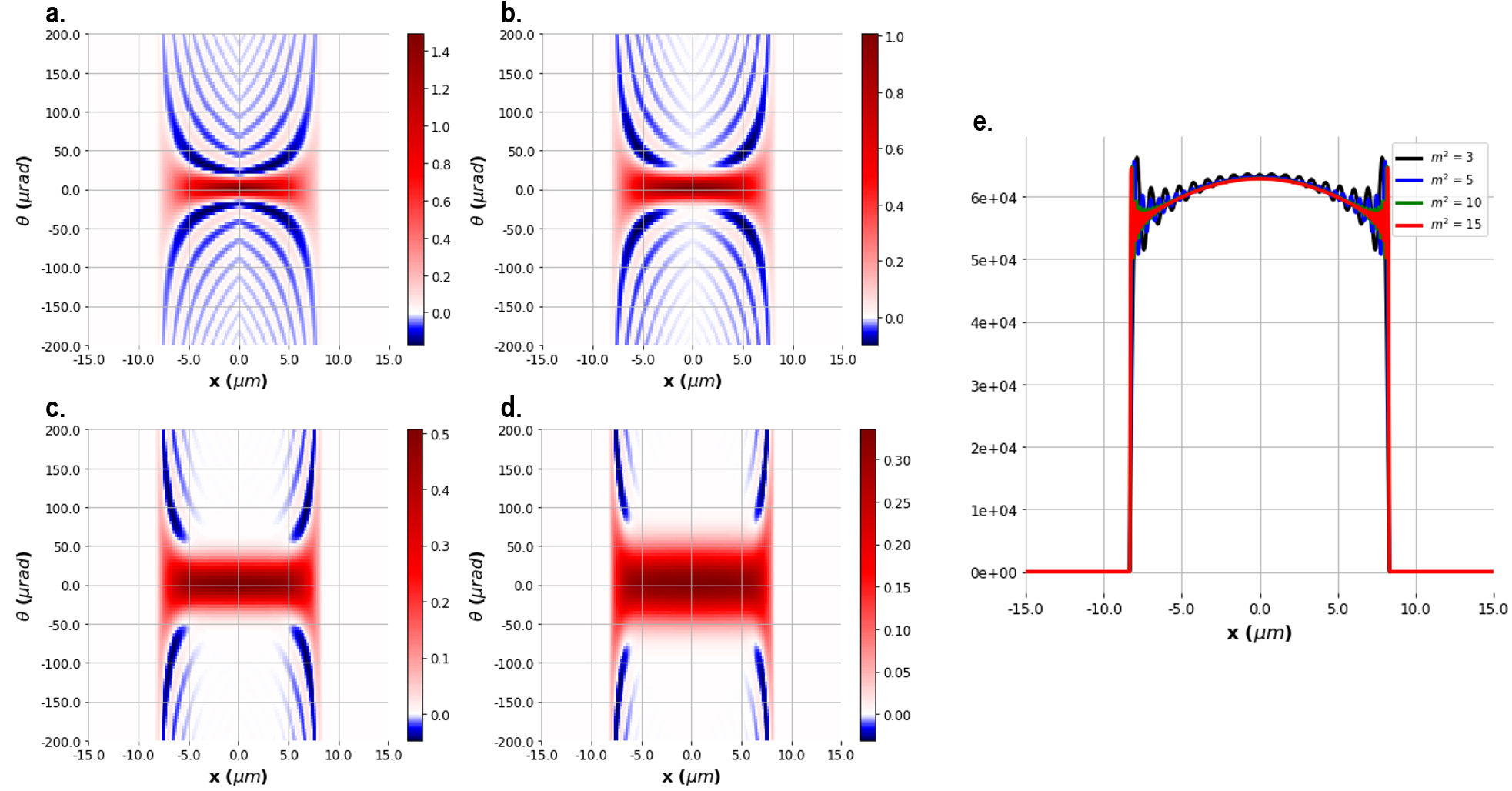}}
\caption{WFD plots for Gaussian beam after passing through single slit aperture.  Beam size, $\sigma_x$, has been fixed and divergence varies according to the parameter $m^2$ in Eqn. (\ref{diffracLimEqn}). The radiation wavelength $\lambda = 3.98$ \AA. (a) $m^2$ = 3, $\sigma_{\theta}$ = 5.90 $\mu$rad (b) $m^2$ = 5, $\sigma_{\theta}$ = 9.83 $\mu$rad (c) $m^2$ = 10, $\sigma_{\theta}$ = 19.66 $\mu$rad (d) $m^2$ = 15, $\sigma_{\theta}$ = 29.49 $\mu$rad (e) projection on spatial axis for cases (a) - (d).}
\label{gsn_expr_Mrange}
\end{figure*}

Figure \ref{gsn_expr_Mrange} shows the WFDs resulting from the convolution of GR with the aperture Wigner filter function displayed in Figure \ref{sftransmission_sfwigner} (b).  This calculation was performed on a discrete grid of size $n_x = 4000$ by $n_{\theta} = 4000$.  Because the analytic expressions are available in the GR case, we are free to increase the resolution to a very fine level without incurring excessively large runtimes and memory demand.  These results demonstrate the diminishing coherence effects for a fixed beam size and increasing divergences corresponding to increasing $m$ values.

%In addition to the analytic calculation, we have also constructed the Wigner function numerically and subsequently applied the convolution with the single slit aperture WFD and performed th

\begin{figure*}[!tbh]
	\centering
{\includegraphics[width=\textwidth,height=3.2in,keepaspectratio]{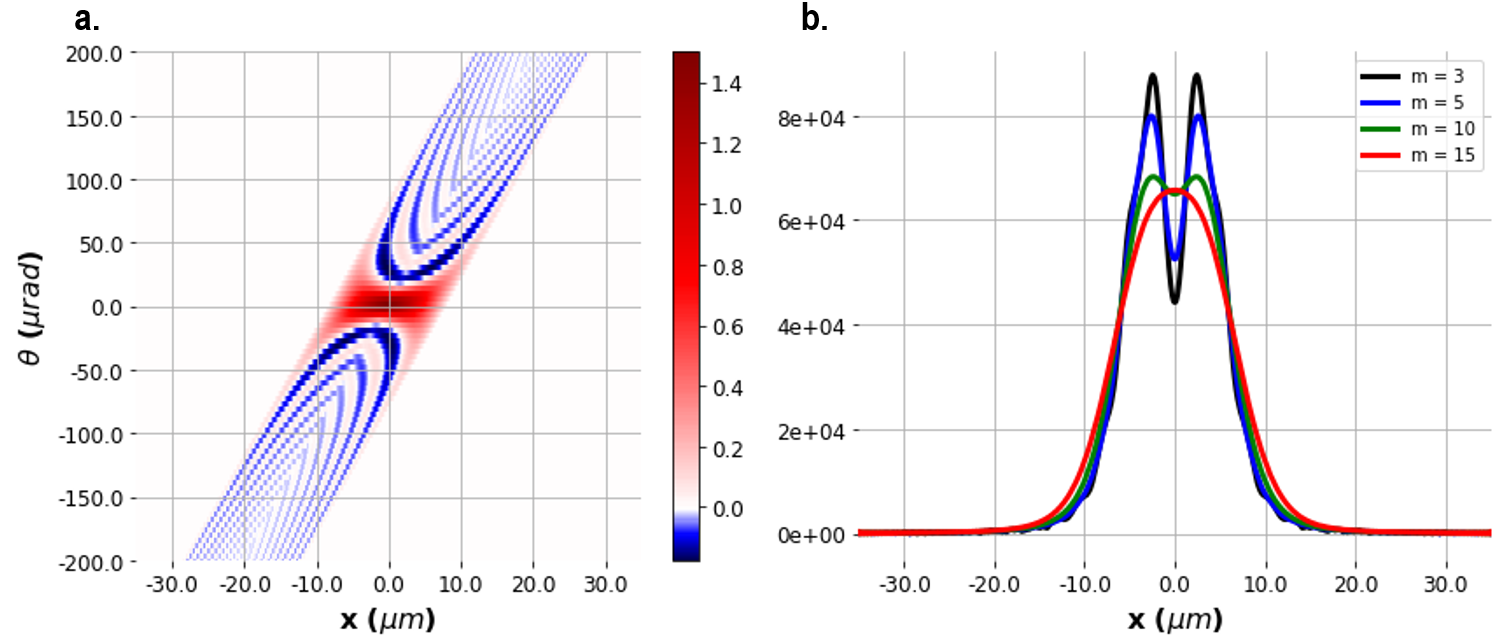}}
\caption{(a) WFD from Figure \ref{gsn_expr_Mrange} for $m^2$ = 3 after drifting 10 cm. (b) Spatial projections of drifted WFDs for $m^2$ = 3, 5, 10 and 15.}
\label{gsn_expr_mrange_drift}
\end{figure*}

\begin{figure*}[!tbh]
	\centering
{\includegraphics[width=\textwidth,height=2.2in,keepaspectratio]{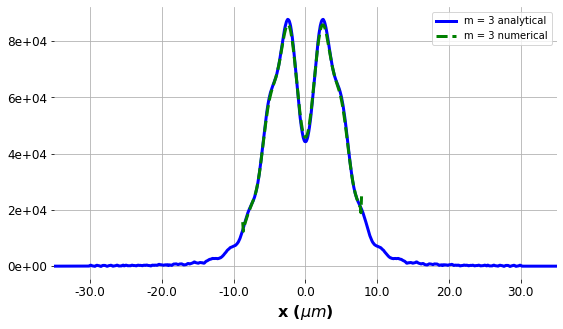}}
\caption{Comparison of analytically and numerically calculated spatial projections of drifted WFDs for m = 3.}
\label{gsn_expr_numeric_m3}
\end{figure*}

Figure \ref{gsn_expr_mrange_drift} shows the intuitive strength of the WFD method by illustrating the mechanism of diffraction in which the oscillations in $\theta$ give rise to interference effects following the drift.  Before the drift, the oscillations of the Wigner functions in $\theta$ will cancel when performing a projection, but following the drift, they result in a large dip at the center of the intensity distribution which can be seen in Figure \ref{gsn_expr_mrange_drift} (b).  In Figure \ref{gsn_expr_mrange_drift} (b), results are shown for increasing values of $m^2$, showing how this interference effect disappears with decreasing coherence.  In order to demonstrate the validity of our numerical computations, Figure \ref{gsn_expr_numeric_m3} compares the spatial projections of the drifted GR WFD at $m^2 = 3$ calculated numerically and analytically.  Because the numerical result must be redeposited onto the initial grid, one tends to truncate the calculation within a smaller range.  The WFD was computed numerically by first constructing a discrete Gaussian along with discrete representation of the single slit aperture WFD.  These two functions were then convolved in $\theta$ to compute the result following the aperture.  Finally, this WFD was drifted for 10 cm by application of the drift transfer matrix and redeposition upon the original grid using the method as first reported in \cite{icap_paper}.

\begin{figure*}[!tbh]
	\centering
{\includegraphics[width=\textwidth,height=2.2in,keepaspectratio]{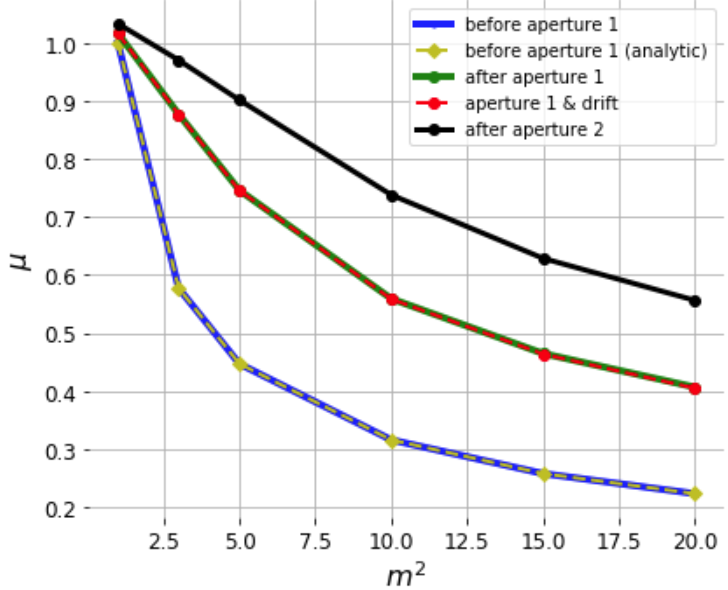}}
\caption{Degree of coherence with varying m values before and after the aperture.}
\label{deg_coherence_gsn01}
\end{figure*}

Figure \ref{deg_coherence_gsn01} displays the dependence of degree of coherence on increasing $m$ value both before and after the single slit aperture.  We show agreement between numerical and analytical (Eqn. (\ref{degCoherenceExpr})) calculations for this quantity before the aperture.  We note that following propagation through the aperture, the degree of coherence has increased which is to be expected since a more coherent subset of the radiation has been selected by the aperture.  In addition, we confirm that the degree of coherence is invariant under the final free space propagation which is predicted by Eqn. (\ref{degCoherenceLinear}).

\section{Examples -- Undulator radiation}
%% !TEX root = ../beamlineMapPrab.tex

\subsection{Partially coherent undulator radiation}

As an example of transporting a partially coherent Wigner function, we will consider the case of undulator radiation resulting from a beam
of electrons in a synchrotron light source.

A single electron with initial phase space coordinates $\vec z_0$ passing through the undulator will produce a coherent wavefront $E_u(x)$.  The electron will be drawn from a distribution of electrons, $f_e(\vec z_0)$, that are circulating in the electron storage ring.  This distribution generally takes on a Gaussian form resulting from an equilibrium between the damping and diffusion effects from synchrotron radiation \cite{Nash2006Mar, Sands1968}.  Due to differences
in the longitudinal coordinates of the emitting electrons, the radiated wavefronts will add incoherently and produce partially coherent radiation.  

We assume that the radiation will satisfy 
\begin{equation}
W(\vec z;\vec z_0) = \wigmap(E(\vec x;\vec z_0)) = W_0(\vec z - \vec z_0)
\end{equation}
where
\begin{equation}
W_0(\vec z) = \wigmap(E(\vec x;\vec z_0 = \vec 0))
\end{equation}
Under these conditions, the multi-electron Wigner function $W_{me}$ may be related to the single electron Wigner function via convolution with the electron beam
distribution:
\begin{equation}\label{WelecDist}
W_{me}(\vec z) = W_{se}(\vec z) \conv f_e(\vec z)
\end{equation}

Next we note that the convolution of the single electron Wigner function with the electron beam distribution may be postponed until the first aperture by applying the following identity.

\begin{equation}
U_{M_1}W_{se}(\vec{z})\conv f_e(\vec{z}) = W_{se}(M_1 \vec{z}) \conv f_e(\vec{z})
\end{equation}
where the transfer matrix, $M_1$, represents the propagation from the source to the first aperture.  Recall the discussion in section II-D for the definition of the operator $U_{M_1}$. This identity is known as the ``emittance convolution theorem'' attributed to K.-J. Kim 
(see e.g. discussion in \cite{Glass2017Jun} and references therein). This theorem allows us to use coherent optics until the first aperture where we then need to construct the partially coherent Wigner function via convolution with the electron beam phase space distribution
that has been propagated to the same position via the transfer matrix of the first section $M_1$. The fact that apertures are represented by only a partial convolution prevents us from extending this identity and applying the convolution beyond the first aperture.

\begin{figure*}[!tbh]
	\centering
{\includegraphics[width=\textwidth,height=3.2in,keepaspectratio]{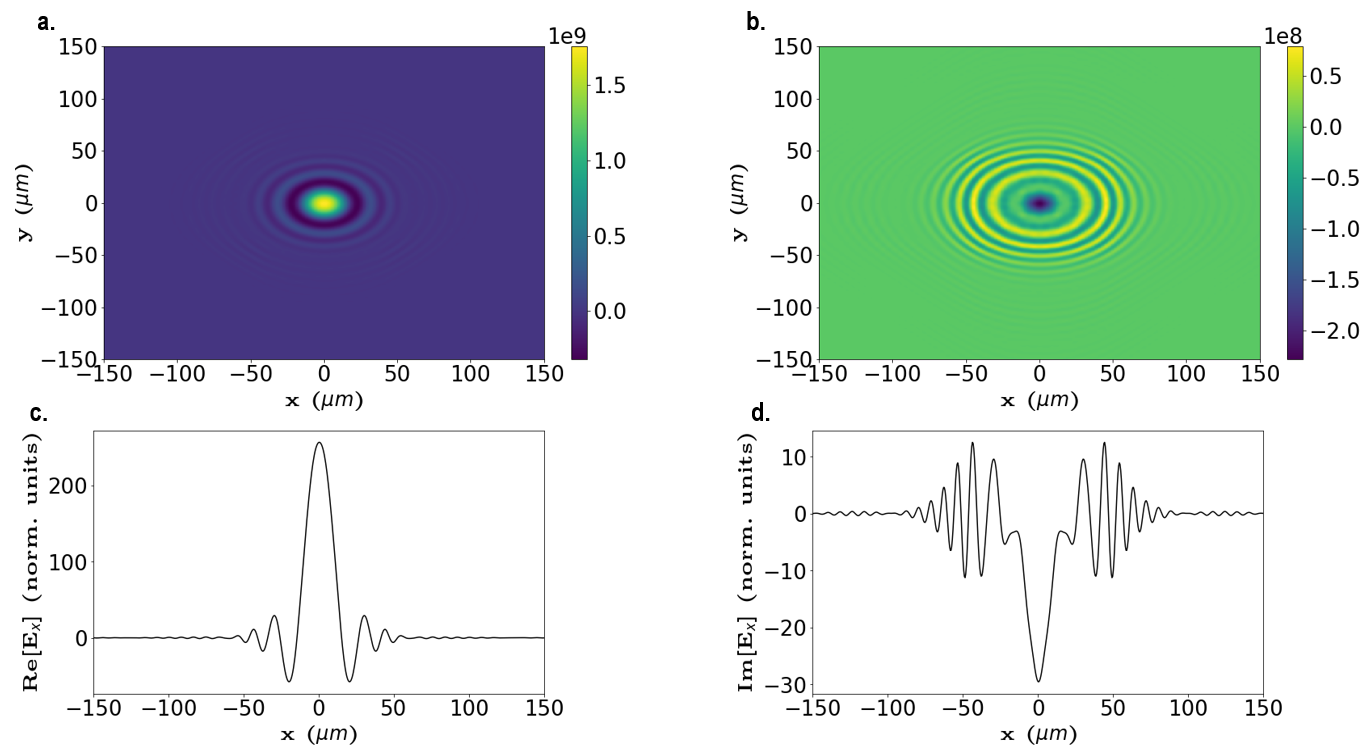}}
\caption{Undulator radiation at first harmonic energy of 3.115 keV. (a) Real part of electric field. (b) Imaginary part of electric field. (c) Horizontal projection of real electric field. (d) Horizontal projection of imaginary electric field.}
\label{initialElecFieldsSrw}
\end{figure*}

As an example of UR, we consider the undulator and electron beam with parameters defined in Table \ref{paramTable01}.  The software package Synchrotron Radiation Workshop (SRW) \cite{Chubar2013}\footnote{The SRW source code is maintained by Oleg Chubar and is avaialble at https://github.com/ochubar/srw} is used for the initial wavefront calculation that will be used to construct the corresponding WFD.  We compute the radiation at the first optical element and progress it through the lens and drift such that we achieve the one-to-one focusing yielding the radiation as it would appear at the center of the undulator.  Figure \ref{initialElecFieldsSrw} (a) and (b)  display the real and imaginary parts of the electric field and in (c) and (d) we have projected these fields onto the horizontal axis and normalized them according to Eqn. (\ref{enormcoord}).  Note that since our planar undulator has a vertical magnetic field, we have selected the dominant polarization component which is horizontal.

\begin{figure*}[!tbh]
	\centering
{\includegraphics[width=\textwidth,height=3.2in,keepaspectratio]{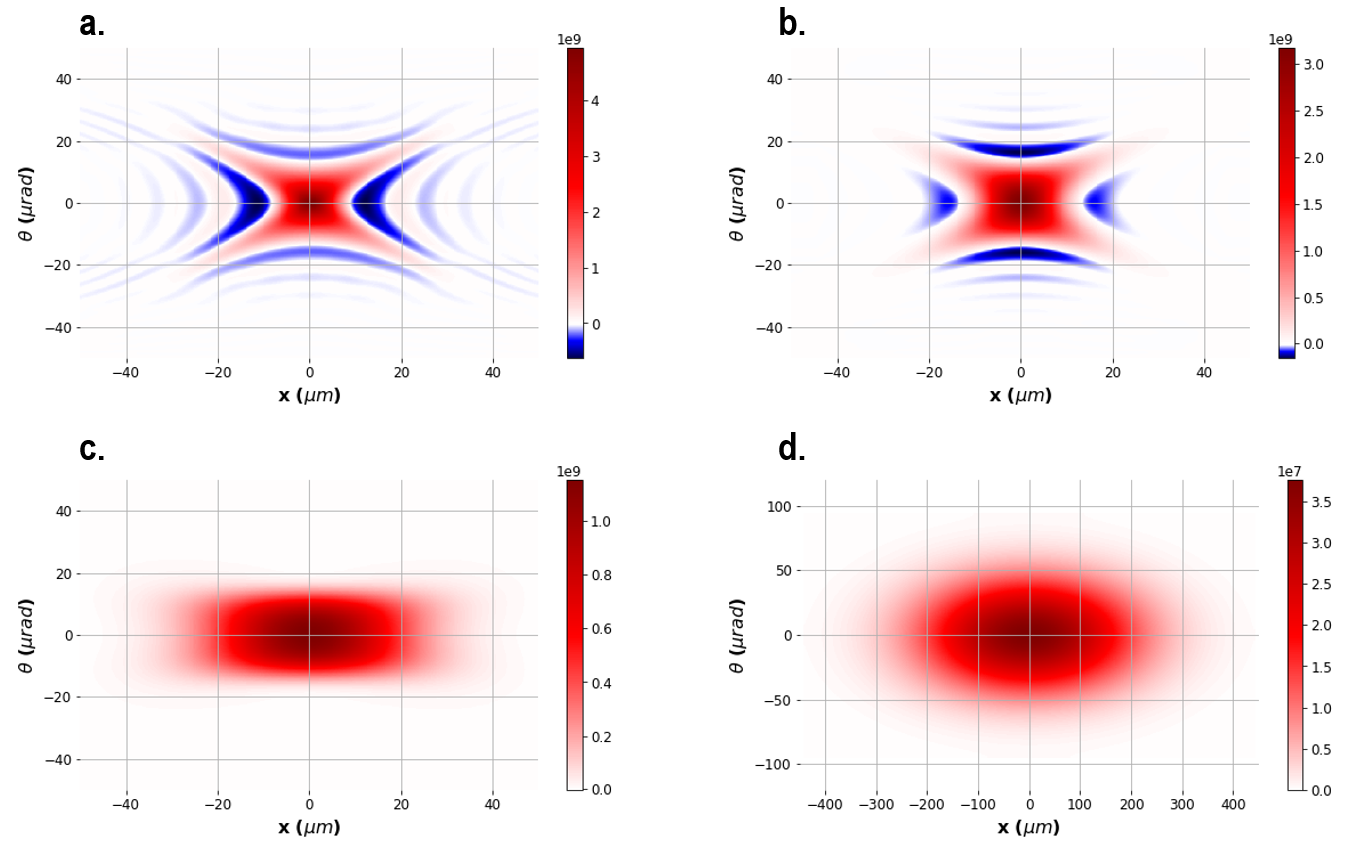}}
\caption{Undulator radiation multi-electron WFD for varying emittance $\epsilon$: (a) zero emittance (fully coherent)  (b) 4 pm  (c) 40 pm  (d) 4 nm.}
\label{ur_fe_conv}
\end{figure*}

The UR WFD is constructed from the electric field projections and is displayed in Figure \ref{ur_fe_conv} (a). Figure \ref{ur_fe_conv} (b) displays the electron beam distribution corresponding to the APS Upgrade parameters given in Table \ref{paramTable01}.  The convolution of the single electron UR WFD and the electron beam distribution results in the multi-electron WFD given in Figure \ref{ur_fe_conv} (c).  In this case, we can see that we are near the diffraction limit as the electron beam size is of the same order as the UR single electron radiation beam size.  The finite emittance result shows properties of both the underlying UR WFD and the Gaussian electron beam distribution\footnote{We note that this result is dependent on the undulator radiation energy used of 3.115 keV. Higher energy radiation will take up a smaller phase space footprint, and the electron beam contribution
will be correspondingly larger. It is thus easier to reach the diffraction limit of high coherence for lower energy radiation, for a fixed electron beam size and divergence.}. This WFD is then propagated through the example beamline and the results from both our WFD transport method and SRW are detailed in the following section.

\subsection{Comparison with multi-electron SRW simulation}

%\begin{figure*}[!tbh]
%	\centering
%{\includegraphics[width=\textwidth,height=2.2in,keepaspectratio]{figures/ur_partcoherent_srw_benchmark01.PNG}}
%\caption{Comparison between WFD method and SRW calculation for partially coherent UR.}
%\label{ur_partcoherent_srw_benchmark01}
%\end{figure*} 

We now report the results from transporting the UR through the single slit aperture and subsequent drift.  The multi-electron WFD is computed by convolution with the electron beam and propagation through the single slit aperture is performed by means of convolution in $\theta$ with the aperture WFD.  The free space drift is performed as done in the Gaussian case with the same linear transport algorithm used in the preceeding section.  We note that this transport algorithm may be applied to any WFD.

This numerical result was benchmarked against a partially coherent SRW calculation wherein the same example beamline was used.  The multi-electron SRW calculation requires running the coherent calculation many times for different macro-electrons.  
In this case of 42pm emittance, we found we could achieve convergence with 5000 macro-particles. Larger emittance, such as 4nm requires larger numbers of macro-electrons: up to 50,000.  The results shown here for 42pm used the more conservative value of 50,000 macro-electrons, though this was not strictly required.

\begin{figure*}[!tbh]
	\centering
{\includegraphics[width=\textwidth,height=2.2in,keepaspectratio]{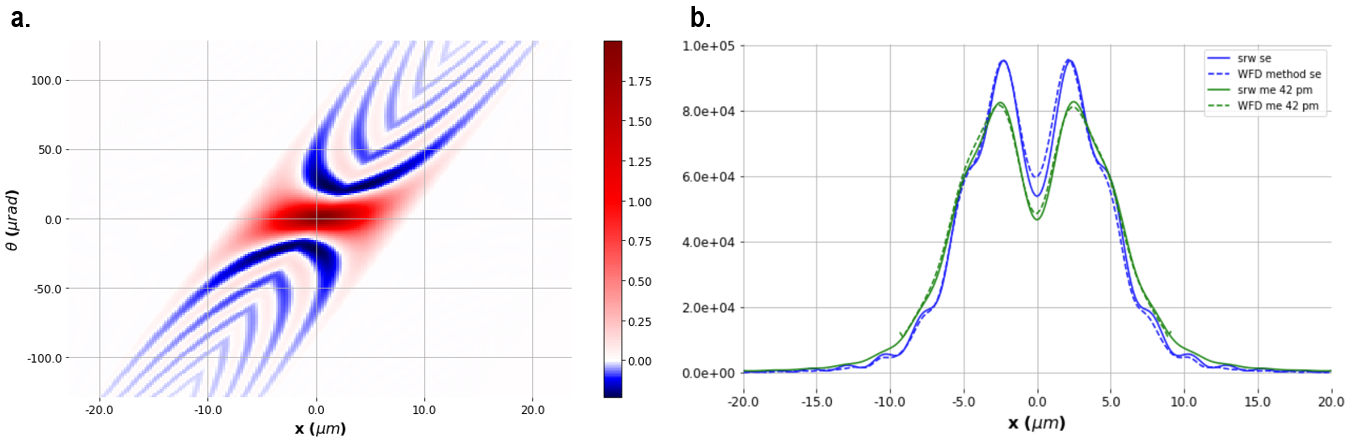}}
\caption{(a) Fully coherent UR WFD following propagation through single slit aperture and 10 cm drift. (b) Comparison between WFD method and SRW calculation for fully coherent, single electron (se), and partially coherent, multi-electron (me), UR spatial projections.}
\label{wfd_srw_se_me_dist_benchmark01}
\end{figure*}

\begin{figure*}[!tbh]
	\centering
{\includegraphics[width=\textwidth,height=2.2in,keepaspectratio]{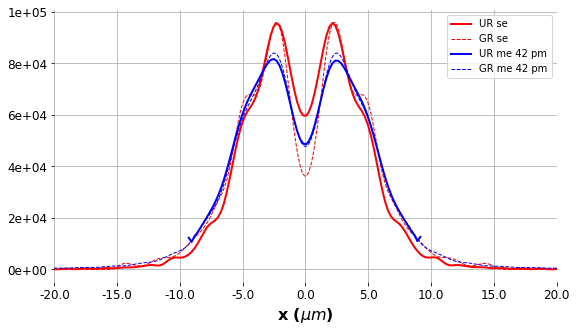}}
\caption{Spatial projections of Gaussian and undulator radiation propagated through single slit aperture and 10 cm drift via WFD method.  In the legend, se denotes single electron (fully coherent) and me denotes multi-electron (partially coherent).  For the latter, the photon beam has been convolved with the APS-U electron beam given in Table \ref{paramTable01}.}
\label{gsn_ur_comparison01}
\end{figure*}

These results are shown in Figure \ref{wfd_srw_se_me_dist_benchmark01}.  Figure \ref{wfd_srw_se_me_dist_benchmark01} (a) displays the final numerically computed WFD.  We note similarity to the Gaussian case being dominated by the structure of the single slit aperture Wigner filter function.  
In Figure \ref{wfd_srw_se_me_dist_benchmark01} (b), the spatial projections of the final WFD have been calculated and used to benchmark our numerical method against the computationally intensive multi-electron SRW calculation.  The partially coherence reflects the effect of the electron 
beam distribution which, as mentioned, has been drawn from the APS Upgrade beam parameters.  We note that the 42.2 pm emmittance electron beam, if representing directly a Gaussian WFD, would correspond to an $m^2$ value of 1.33.  Recall Eqn. (\ref{diffracLimEqn}). Figure \ref{gsn_ur_comparison01} provides a 
comparison between the propagated Gaussian and Undulator radiation for the cases of fully and partially coherent radiation. For the Gaussian radiation, the second moments have been taken from adding in quadrature the undulator $\sigma_x$ and $\sigma_\theta$ with those of the electron
beam. We note that for the fully coherent case (single electron), there is not perfect agreement between the two calculations, with the Gaussain case having a larger central dip from diffraction. In the case of the 42 pm emittance, however, the Gaussian result quite adequately 
reproduces the more complex undulator radiation calculation.

\section{Conclusion}
%% !TEX root = ../beamlineMapPrab.tex

We have described a new, unified method for transporting coherent and partially coherent radiation through a beamline of linear optical elements and apertures. This approach relies on transporting the Wigner distribution of the radiation wavefront in a manner akin to single-particle tracking in particle accelerators. In contrast with physical optics modeling, the formalism is the same for fully coherent and partially coherent radiation, and has the same computational time and complexity. The matrices in the matrix-aperture beamline we have defined may be computed with ray tracing, 
in the general case, or analytically for simplified beamline models. We thus show how ray tracing may be unified with wave optics, at least within this linear approximation. 

We have demonstrated this approach by transporting both a gaussian wave front and an undulator radiation wavefront through drift space and a single slit aperture. In the case of the gaussian wave front, we find excellent agreement between an analytical result, the Wigner distribution approach, and a the well-established SRW physical optics code. For the undulator radiation, we have excellent agreement between the Wigner distribution approach and multi-electron SRW calculations. Finally, we have seen that for large enough emittance (even as small as 42pm horizontal emittance), the Gaussian
results approximate the undulator radiation to a high degree of accuracy.

Future work will extend this technique to nonlinear optical elements, which will allow us to account for optical aberrations. We have briefly described the formalism for nonlinear optical elements in Appendix~\ref{sec:nl_maps}.
Extension to the non-linear case will raise new issues, for future exploration. See e.g. \cite{Dragt1998Jun} for some understanding
of the potential resulting complications.

\section{Acknowledgments}
The authors would like to thank A. Wojdyla for valuable discussion and for alerting us to the literature on Linear Canonical Transforms, including reference \cite{Healy_2016}.
This material is based upon work supported by the U.S. Department of Energy, Office of Science, Office of Basic Energy Sciences, under Award Numbers DE-SC0020593 and DE-SC0018571.

\begin{appendix}

\section{Coherent Wave Optics}\label{app:cwo}
%% !TEX root = ../beamlineMapPrab.tex

Here we describe, for completeness, the approach used for propagating a radiation wavefront through a beamline using physical optics techniques. The propagation of this wavefront through an optical beamline is the goal of physical optics software such as SRW and XRT \cite{XRTdoc, Chernikov2017Sep}.
In these codes, there exist a variety of numerical implementations of propagators to transport the wavefront through free space, lenses, mirrors, gratings, and other optical elements.

We begin with a complex electric field $E_0(x,y,s=0)$ at the entrance of the beamline.  Optical transport beamlines in synchrotron radiation facilities can often be represented with the scalar paraxial optics approximation~\cite{Bahrdt1997Nov,Chubar2002Sep}. For example, the near-field Fresnel integral for free-space propagation can be written as a convolution:
\begin{equation}
E(x,y,s) = E_0(x,y) \conv h(x,y,s)
\end{equation}
with
\begin{equation}
h(x,y,s) = \frac{e^{i k s}}{i\lambda s} e^{i\frac{k}{2z}(x^2+y^2)}
\end{equation}
A thin lens may be traversed via
\begin{equation}
E(x,y,s') = e^{-i\frac{k}{2f}(x^2+y^2)}E(x,y,s)
\end{equation}
with $k=\frac{2\pi}{\lambda}$, $n$ is the index of refraction, and $f$ is the focal length~\cite{Goodman2005}. 

The combination of drifts, lenses, and focussing mirrors together can be combined to create
a symplectic transport matrix $M$ for the geometric ray optics as given in Eqn. (\ref{TransferMatrix}). Knowing $M(s)$, one may propagate the wavefront through the linear beamline down the channel via Linear Canonical Transformation (LCT)~\cite{Healy_2016}. Explicitly for 4D phase space, the transfer matrix is written in the form
\begin{equation}
M = \begin{pmatrix}
A & B\\
C & D
\end{pmatrix}
\end{equation}
The transformed electric field $E_f(\vec x)$ is given by
\begin{equation}
E_f(\vec x) = \frac{1}{\sqrt{\det(iB)}} \int e^{i\pi \vec u^T \mathcal{M} \vec u} E_i(\vec x) d\vec x_i
\end{equation}
with
\begin{equation}
\vec u = \begin{pmatrix}
\vec x_f& \vec x_i
\end{pmatrix}
\end{equation}
and
\begin{equation}
\mathcal{M} =  \begin{pmatrix}
DB^{-1} & -B^{-1}\\
-B^{-1} & B^{-1} A
\end{pmatrix}
\end{equation}
where the subscripts $f$ and $i$ represent initial and final.

We point out here, that in addition to use of analytical expressions for beamline elements to determine the Hamiltonian, and thus find the transfer matrix $M(s)$, one may also use a ray tracing code, set up the beamline, and by tracking a series of rays offset from the central trajectory, 
derive the transport matrix numerically along the beamline.

The effect of the apertures, represented by the transfer functions $t_j(x,y)$ on the wavefront simply by multiplication:
\begin{equation}
E(x,y;s') = t_j(x,y)E(x,y;s)
\end{equation}

\section{Normalization of wavefronts and Wigner functions}
In this paper, we will assume electric fields which satisfy the separability condition
\begin{equation} \label{sepcond}
E(x,y;s) = E_0 E_x(x;s) E_y(y;s),
\end{equation}
where $E_0$ is a constant with units of electric field.

%% !TEX root = ../beamlineMapPrab.tex

We normalize the separate electric field components in 1-D such that 
\begin{eqnarray}\label{enormcoord}
\int_{-\infty}^{\infty}E^{*}(x)E(x)dx=1,\\ 
\int_{-\infty}^{\infty}E^{*}(\theta)E(\theta)d\theta=1.
\end{eqnarray}

Following Bazarov~\cite{Bazarov}, we have normalized the electric field in the same way as wave functions are normalized in quantum mechanics.  The second moments of the field distribution in coordinate and angular representations may now be calculated as
\begin{eqnarray}
<x^2> &=& \int_{-\infty}^{\infty} x^2 E^*(x)  E(x) dx,\\
<\theta^2> &=& \int_{-\infty}^{\infty} \theta^2 E^*(\theta) E(\theta) d\theta.
\end{eqnarray}

We now introduce the Wigner function defined from the electric field, $E(x)$, as follows
\begin{equation}
W(x,\theta) = \frac{1}{\lambda}  \int_{-\infty}^{\infty} E^*\left(x-\frac{\phi}{2}\right)E\left(x+\frac{\phi}{2}\right)e^{\frac{-2\pi i}{\lambda}\phi \theta}d\phi,
\end{equation}
where $W(x,\theta)$ will be normalized as
\begin{equation}\label{wignerNorm01}
\int_{-\infty}^{\infty} \int_{-\infty}^{\infty} W(x,\theta)dxd\theta=1.
\end{equation}

The Wigner function can be thought of as a probability distribution in phase space except for the fact that it may become negative. The second moments are given simply as
\begin{eqnarray}
<x^2> &=& \int_{-\infty}^{\infty} \int_{-\infty}^{\infty} x^2 W(x,\theta) dxd\theta,\\
<\theta^2> &=& \int_{-\infty}^{\infty} \int_{-\infty}^{\infty} \theta^2 W(x,\theta) dxd\theta,\\
<x \theta> &=& \int_{-\infty}^{\infty} \int_{-\infty}^{\infty} x \theta W(x,\theta) dxd\theta.
\end{eqnarray}

\section{Transport of Wigner function under non-linear maps}\label{sec:nl_maps}
%% !TEX root=../beamlineMapPrab.tex

For propagation of the Wigner function, Bazarov (Property 5) uses the Hamiltonian $\mathcal{H} = \frac{\hat p^2}{2m} + V(x)$ for the Quantum Mechanics case of a particle in a potential.
We have used a different Hamiltonian for the optics case. The problem of general transport of the Wigner function remains.
The evolution equation for the Wigner function under a general Hamiltonian is given as follows~\cite{Curtright2011Apr}
\begin{equation}
\frac{\partial W(x,\theta_x, y, \theta_y;s)}{\partial s} = [W,H]_*,	
\end{equation}
where the Moyal bracket is defined for arbitrary phase space functions $f$ and $g$ as
\begin{equation}
[f,g]_{*} = \frac{1}{i \lambdabar}(f*g-g*f)
\end{equation}
and the Moyal star is given by
\begin{equation}
* = e^{\frac{i\lambdabar}{2} \left(\overleftarrow\partial_x \overrightarrow\partial_{\theta} - \overleftarrow\partial_{\theta} \overrightarrow\partial_x \right)}
\end{equation}
with the arrows representing action of the derivative, either to the left or right, depending on arrow orientation.

\section{Analytical Calculation of a Gaussian Wigner Distribution Through a Single Slit}

%%!TEX root = ../beamlineMapPrab.tex

The passage of a Gaussian Wigner distribution through a slit can be evaluated in terms of the integral
\begin{equation}
\mathcal{I}(x, \theta) = \int_{-\infty}^{\infty} d\tau ~ \frac{\sin \tau q}{\tau} \exp \left \{ - \frac{(\theta - \tau)^2}{2 \sigma_\theta^2} \right \}
\end{equation}
It is convenient to normalize all the variables to $\sigma_\theta$, so that $\hat{\theta} = \theta/\sigma_\theta$, $\hat{\tau} = \tau/\sigma_\theta$, and $\hat{q} = \sigma_\theta q$, so that the integral becomes:
\begin{equation}
\mathcal{I}(x, \theta) = \int_{-\infty}^{\infty} d\hat{\tau} ~ \frac{\sin \hat{\tau} \hat{q}}{\hat{\tau}} \exp \left \{ - \frac{1}{2}(\hat{\theta} - \hat{\tau})^2 \right \}
\end{equation}
This integral can be rewritten as the imaginary part of an indefinite integral with respect to $q$, 
\begin{equation}
\mathcal{I}(x, \theta) = \textrm{Im} \left [ i \int d\hat{q} \int_{-\infty}^{\infty} d\hat{\tau} ~ e^{i \hat{\tau} \hat{q}} \exp \left \{ - \frac{1}{2}(\hat{\theta} - \hat{\tau})^2 \right \} \right ]
\end{equation}
Expanding the Gaussian argument brings a Gaussian $\theta$ envelope out front:
\begin{equation}
\mathcal{I}(x, \theta) = e^{-\frac{\hat{\theta}^2}{2}} \textrm{Im} \left [ i \int d\hat{q} \int_{-\infty}^{\infty} d\hat{\tau} ~ \exp \left \{ - \frac{1}{2}(\hat{\tau}^2 + 2 i \hat{\tau} (\hat{q} + i \hat{\theta})) \right \} \right ].
\end{equation}
The argument can be simplified by completing the square, noting that
\begin{equation}
\hat{\tau}^2 + 2 (\hat{\theta} + i \hat{q}) \hat{\tau} = (\hat{\tau} + \hat{\theta} + i \hat{q})^2 - (\hat \theta + i \hat q)^2
\end{equation}
which then gives the integral as
\begin{equation}
\mathcal{I}(x, \theta) = e^{-\frac{\hat{\theta}^2}{2}} \textrm{Im} \biggl [ i \int d\hat{q} e^{\frac{1}{2} (\hat\theta + i \hat q)^2} \underbrace{\int_{-\infty}^{\infty} d\hat{\tau} ~ \exp \left \{ - \frac{1}{2}(\hat{\tau}^2 + i\hat{q} + \hat{\theta}) \right \}}_{\sqrt{2 \pi}} \biggr ]
\end{equation}
and the integral becomes
\begin{equation}
\mathcal{I}(x, \theta) = \sqrt{2 \pi}e^{-\frac{\hat{\theta}^2}{2}} \textrm{Im} \biggl [ i \int d\hat{q} ~ e^{\frac{1}{2} (\hat\theta + i \hat q)^2} \biggr ]
\end{equation}
and which is then given by
\begin{equation}
\mathcal{I}(x, \theta) = \sqrt{\frac{\pi}{2}} \textrm{Im} \biggl [ i ~\textrm{erf} \left ( \frac{\hat{q} + i \hat{\theta}}{\sqrt{2}} \right )\biggr ]
\end{equation}

\end{appendix}

% Add 'References' to the Table of Contents
%\addcontentsline{toc}{section}{References}
% Create bibliography at the end of the  of the document
%\printbibliography
\bibliography{beamlineMapPrab}

\end{document}